# Methods for Uncertainty Representation in Risk Management: A Comparative Review and Decision-Oriented Framework

**Albert Kutej[a] , Stefan Rass[b]**

[a] University of Klagenfurt, Universitätsstraße 65-67, 9020 Klagenfurt am Wörthersee, Austria, e-mail: akutej@edu.aau.at, ORCID: https://orcid.org/0009-0001-5952-3338
[b] LIT Secure and Correct Systems Lab, Johannes Kepler University Linz, Altenberger Straße 69, 4040 Linz, Austria, e-mail: stefan.rass@jku.at, ORCID: https://orcid.org/0000-0003-2821-2489

## Abstract

The consideration of uncertainty is a central but frequently inadequately addressed component of risk management. A systematic treatment of uncertainty is essential for ensuring the quality and traceability of decision-making processes, particularly in complex and safety-critical environments. This review systematically analyzes how established risk management approaches conceptualize and represent uncertainty in both their theoretical foundations and practical applications. Based on a systematic literature review of 370 publications, the identified approaches are classified into five methodological families. These include probabilistic methods, evidence-based and fuzzy-logic approaches, qualitative elicitation techniques, graphical and visual representations and hybrid frameworks. The analysis shows that probabilistic methods remain predominant due to their quantitative rigor, whereas fuzzy and evidence-based approaches are particularly suited to addressing vagueness and epistemic uncertainty. Qualitative and graphical approaches are found to enhance interpretive understanding and support the transparent communication of uncertainty. Despite these developments, the analysis indicates that the practical integration of these approaches into operational risk management remains limited in many domains. The findings highlight the need for more structured guidance in method selection and suggest that future research would benefit from further development of hybrid approaches and visualization techniques.

# 1. Introduction

Uncertainty is a fundamental component of risk and influences how risks are identified, assessed and communicated across disciplinary contexts [1]. Although uncertainty is an essential element of risk assessment, it is frequently addressed only implicitly or reduced to simplified probability estimates that fail to capture the full range of underlying knowledge gaps [2,3]. Recent research emphasizes the need to conceptualize and treat uncertainty systematically in order to strengthen the transparency, interpretability and robustness of risk analyses [4]. This review examines how different approaches [5–15] conceptualize, capture, model and communicate uncertainty within current risk assessment practice. It synthesizes central theoretical foundations and relevant methodological developments and compares probabilistic, evidential, fuzzy, qualitative, graphical and hybrid approaches based on representative contributions from the literature. The analysis illustrates the different forms of uncertainty in complex systems and provides a structured overview of options that address different forms of epistemic and aleatory uncertainty. The following sections outline the problem statement, define the review objectives and analytical scope and summarize the methodology for identifying and selecting relevant literature.

## 1.1. Background and Problem Statement

In risk assessment, the consideration of uncertainty plays a significant role because it influences how risks are identified, interpreted and managed across a wide range of application domains. Risk assessments increasingly confront complex system behaviors, heterogeneous data sources and dynamic environments in which uncertainty affects both the estimation of likelihoods and the evaluation of potential consequences [16]. As system interconnectivity grows and decisions must be made under conditions of incomplete or unreliable information, the explicit treatment of uncertainty becomes a fundamental prerequisite for analytical transparency and decision quality [17,18]. In this context, uncertainty is not a marginal aspect of risk assessment but a central dimension that determines the validity and interpretability of any risk judgment. Despite this relevance, uncertainty is treated only implicitly in many established assessment methods or is reduced to single probability estimates that neither reflect possible ranges nor make underlying assumptions visible and therefore fail to capture existing knowledge limitations adequately [2]. Conceptual ambiguities concerning the meaning and scope of uncertainty persist and distinctions between variability, ambiguity, epistemic uncertainty and aleatory uncertainty are addressed in divergent ways across frameworks and disciplines. Such conceptual inconsistency has direct practical implications. It can lead to methodological misalignment, underestimation of knowledge gaps and insufficient reflection on assumptions that influence the outcomes of risk assessments [19]. As a result, uncertainty remains insufficiently and inconsistently addressed in many risk analysis practices, even though it substantially affects the robustness, transparency and traceability of assessment outcomes. The central research problem addressed in this review is that uncertainty, despite being a defining characteristic of risk [1], is not systematically or consistently treated across existing assessment approaches. Current methods differ substantially in how uncertainty is conceptualized, represented, and operationalized, resulting in a fragmented landscape of possibilities. At the same time, structured and practically applicable guidance for selecting appropriate uncertainty representation methods in applied risk management contexts remains limited. As a result, the choice for a method is often made implicitly or without transparent justification, which can affect the interpretability and robustness of risk assessments. Motivated by this, we examine how different frameworks define uncertainty, how they integrate it into assessment procedures, and which conceptual and theoretical advancements have been proposed for a systematic treatment of uncertainty for risk management. Several review studies have already addressed uncertainty in risk analysis and risk management, including the work of Zio and Pedroni [1] and Aven and Zio [20], which provide important conceptual foundations for understanding uncertainty and its role in risk-related decision-making. However, these contributions primarily focus on conceptual distinctions and general frameworks, with less emphasis on systematic comparisons of approaches for uncertainty representation in applied risk assessment. Given these challenges, the present review contributes to the literature in three main ways. First, it provides a structured classification of uncertainty representation methods into five methodological families, enabling a systematic comparison across approaches that are often treated separately. Second, it emphasizes practical applicability by proposing a decision-oriented pathway that supports method selection in practical risk management settings. Third, it explicitly addresses different forms of uncertainty, including ambiguity and ignorance, and discusses their implications for method selection.

## 1.2. Objective of the Review

The objective of this review is to provide a structured and comparative overview of methods for addressing uncertainty in risk assessment, with a particular focus on the classification of major method families, the identification of their underlying assumptions, and the analysis of their capacities to represent different forms of uncertainty. The review further aims to support method selection in applied contexts by linking conceptual distinctions of uncertainty to practical decision requirements. By integrating and comparing existing contributions, the review seeks to strengthen conceptual clarity and highlight developments that can improve the treatment of uncertainty in risk analysis.

## 1.3. Analytical Focus of the Review

The analytical focus of this review is structured around four thematic areas that guide the examination of uncertainty across the methodological categories considered. The analysis first examines how uncertainty is conceptually defined within existing frameworks and how distinctions such as epistemic and aleatory uncertainty shape assumptions. It then reviews the spectrum of methods that have been developed to capture or represent uncertainty, including qualitative, quantitative and hybrid techniques. Building on this, the review analyses how these approaches differ in their capacity to represent various dimensions of uncertainty and evaluates their respective strengths and limitations in practical application. Finally, it explores how different representations of uncertainty influence risk interpretation, decision making and the communication of risk-related information. Together, these thematic areas provide the analytical structure for the comparative assessment presented in the subsequent sections.

## 1.4. Scope and Delimitation

The scope of this review encompasses frameworks and methods for identifying, representing and analyzing uncertainty in risk assessment. Rather than centering on individual tools or application specific techniques, the review examines broader groups of methods and their underlying principles. It includes contributions from risk analysis that address contexts in which uncertainty plays a central analytical role. The literature base consists of foundational works as well as recent contributions in order to reflect both established concepts and current theoretical developments. The review excludes purely normative discussions on risk ethics and technical safety standards unless they explicitly address the treatment of uncertainty. These delimitations ensure that the analysis remains focused on methodological aspects that directly support the conceptualization and operationalization of uncertainty within risk assessment.

## 1.5. Systematic Literature Review

The literature considered in this review was identified through structured searches in Scopus and Web of Science, covering major publishers and conference venues (e.g., IEEE, ACM, Springer, Wiley, Elsevier). The search strategy combined terms such as "uncertainty", "risk management", "assessment", and "methods" using varying logical connectives, including both keyword and phrase searches, to systematically capture contributions. Exploratory searches in Google Scholar, including direct and keyword-variant queries, were conducted to complement the database search and to include potentially non-indexed sources. Across all retrieval channels, approximately 370 candidate publications were collected and managed in a dedicated Zotero library. Following deduplication and screening, only studies meeting the inclusion criteria were retained for final analysis. Inclusion required (i) explicit relevance to uncertainty representation in risk assessment or risk management, (ii) independent verification of bibliographic details through established databases or publisher libraries, and (iii) full-text access to enable content validation. Studies lacking topical relevance or sufficient verifiability were excluded. Foundational theoretical works were considered alongside recent contributions to reflect both established concepts and current developments. The selection followed an interdisciplinary orientation and incorporated relevant work from risk analysis, decision science, systems engineering, and related fields. This strategy provides a reproducible and interdisciplinary basis for comparative analysis. The study selection followed a structured multi-stage procedure. First, relevant publications were identified through database searches and complementary exploratory searches. Second, duplicate records were removed and titles and abstracts were screened to assess their relevance to uncertainty representation in risk assessment. Third, full-text screening was conducted based on predefined inclusion and exclusion criteria. The final set of included studies comprises those contributions that met all criteria regarding relevance, verifiability, and accessibility. While the review does not follow a formal PRISMA protocol, the described procedure ensures transparency and supports the reproducibility of the study selection process. In addition, the classification of methods into methodological families was based on their underlying representation logic, including probabilistic, evidential, fuzzy, qualitative, graphical, and hybrid approaches, enabling a consistent comparative framework.

# 2. Theoretical Foundations

This section outlines the conceptual foundations of risk and uncertainty that underpin the subsequent analysis.

## 2.1. Conceptual Foundations of Risk and Uncertainty

A clear conceptual understanding of risk and uncertainty is essential for analyzing how different approaches represent uncertainty in risk management. In established definitions, risk is commonly understood as a function of consequences and their associated likelihoods (often operationalized via probability estimates), reflecting both the severity of potential outcomes and the anticipated chance that these outcomes occur [21,22]. In this review, probability refers to a formal quantitative representation of uncertainty, typically derived from empirical observations or statistical modeling, whereas likelihood is used more broadly to express plausibility when evidence is limited or indirectly available. Uncertainty, in contrast, is interpreted in different ways within the literature. In some approaches, uncertainty is treated as an integral component of risk itself, while in others it is conceptualized as a descriptor of the analyst's knowledge about the system, the available evidence and the adequacy of the underlying models [23,24]. In this review, we adopt the latter perspective, following contributions that distinguish between risk as a function of consequences and likelihoods and uncertainty as a qualifier of the confidence in these estimates. From this perspective, uncertainty affects the credibility and interpretability of any risk estimate. Several distinctions are important for ensuring clarity. Variability refers to real-world fluctuations in system behavior that arise from inherent randomness. In risk assessment, this type of uncertainty is commonly referred to as aleatory uncertainty, and it persists even under complete information [25,26]. Ambiguity, in the context of this review, reflects uncertainty that arises from vagueness, multiple possible interpretations or inconsistent definitions of central concepts and is therefore treated as a specific expression of epistemic uncertainty [16,23]. However, epistemic uncertainty goes beyond ambiguity and includes forms of incomplete, unreliable or conflicting knowledge, such as uncertainty due to data limitations, model misspecifications, parameter variability and uncertainty about causal mechanisms or interactions within the system [27,28]. A third category that is frequently discussed in the literature is ignorance, which refers to situations in which relevant outcomes (e.g., ranges), mechanisms or dependencies are not known or cannot be meaningfully characterized [16,22,29]. While ignorance is sometimes treated as an extreme form of epistemic uncertainty, several authors distinguish it as a qualitatively different condition, since it involves not only incomplete knowledge but also unrecognized or unknowable aspects of the system. In such situations, neither probability distributions nor plausible ranges can be specified in a meaningful way. In this review, ignorance is treated as a distinct analytical category in order to highlight situations where uncertainty extends beyond incomplete knowledge and includes fundamental limits of model specification and system understanding. This distinction is particularly relevant for method selection, as many uncertainty representation techniques implicitly assume that at least partial knowledge can be formalized. These conceptual distinctions, including the differentiation between epistemic uncertainty and ignorance, provide an important foundation for the analysis that follows.

## 2.2. Aleatory and Epistemic Uncertainty

Distinguishing aleatory from epistemic uncertainty is central to the selection and interpretation of methods in risk assessment. The distinction clarifies whether uncertainty arises from inherent randomness or from limitations in knowledge and therefore determines which methods and modeling techniques are appropriate and what the results of risk assessments can legitimately be expected to represent [21,25]. The following subsections provide definitions, causes, illustrative examples, modeling techniques and typical domains of application for each type of uncertainty.

### 2.2.1. Aleatory Uncertainty

Aleatory uncertainty reflects variability inherent in the system, for example fluctuations that occur even under fully known conditions [30]. It reflects the fact that even under perfect knowledge, outcomes cannot be predicted with certainty because they are governed by intrinsic randomness [11,26]. Typical sources include biological variations, environmental fluctuations and stochastic behavior in physical or technical systems. Examples include variations in component lifetimes, fluctuations in patient inflow or random deviations in production processes despite controlled conditions [31]. Aleatory uncertainty is irreducible; additional data may improve the estimation of probability distributions, but the variability itself remains [32]. Probabilistic modeling, through methods such as probability distributions, Monte Carlo simulation, stochastic processes and Bayesian parameter estimation, is the predominant approach to representing aleatory uncertainty [1,13]. These methods form the foundation of reliability engineering, probabilistic risk assessment and quantitative modeling in epidemiology and finance, where the focus lies on quantifying randomness and calculating associated risk metrics.

### 2.2.2. Epistemic Uncertainty

Epistemic uncertainty arises from incomplete knowledge, limited data, measurement errors or insufficient understanding of system structures and mechanisms [23,26]. It reflects the analyst's degree of knowledge rather than intrinsic randomness. Typical sources include missing or sparse evidence, uncertain parameter estimates, ambiguous causal mechanisms or expert disagreement in new or rapidly changing contexts [27]. Unlike aleatory uncertainty, epistemic uncertainty is reducible through improved data collection, model refinement, expert elicitation or validation efforts. Modeling approaches include fuzzy logic, evidence theory, interval analysis, scenario analysis and qualitative elicitation techniques, all of which aim to capture ambiguity, partial belief or incomplete information [33]. Epistemic uncertainty plays a particularly important role in strategic risk analysis, early phases of technological development, crisis management and clinical decision making where empirical evidence is limited or just evolving [34]. A structured comparison of both uncertainty types is presented in **Fehler! Ungültiger Eigenverweis auf Textmarke.**1. As shown in the table, the distinction is not purely conceptual but has practical implications, since methods suited for capturing randomness may inadequately represent incomplete knowledge, while techniques designed to capture epistemic uncertainty may not address stochastic variability [1,21]. Recognizing the interplay between aleatory and epistemic components is particularly important in interconnected domains such as healthcare and critical infrastructures, where uncertainty arises simultaneously from intrinsic system behavior and from informational or conceptual limitations [23,27]. The conceptual distinctions outlined in this chapter provide the analytical foundation for the review that follows. Understanding whether uncertainty arises from inherent variability or from limitations in knowledge is essential for assessing the suitability, strengths and limitations of different approaches [1,16,21,23]. Beyond the approaches covered in this review, several additional frameworks address uncertainty in risk and decision-making. These include imprecise probability theory [35,36], probability bounds analysis (p-boxes), Info-Gap theory [37], and decision-making under deep uncertainty [38] (DMDU), including approaches such as Robust Decision Making. These frameworks provide important perspectives, particularly for situations characterized by severe epistemic uncertainty, limited data, or deeply uncertain system behavior. However, they are only partially aligned with the primary focus of this review, which is on methods commonly applied in operational risk assessment and risk management contexts. Therefore, these frameworks are considered complementary and are not included in the core methodological classification presented in this paper. Building on these conceptual foundations, the next section examines how established families capture, model and represent uncertainty in risk management.

Table 1
Conceptual distinction between aleatory and epistemic uncertainty

| Aspect | Aleatory Uncertainty | Epistemic Uncertainty |
|---|---|---|
| **Definition** | Inherent randomness or natural variability in system behavior that persists even under complete information | Lack of knowledge, lack of precision or incomplete information about parameters, models or mechanisms |
| **Nature** | Objective<br><br>(independent of the analyst's knowledge) | Subjective<br><br>(depends on the quality and completeness of available information) |
| **Reducibility** | Irreducible<br><br>(additional data do not eliminate variability) | Reducible through improved data, research, model refinement or expert elicitation |
| **Typical Representation** | Probability distributions, stochastic models, Monte Carlo simulation, Bayesian inference | Fuzzy sets, evidence theory, interval analysis, expert elicitation, scenario analysis |
| **Sources** | Natural variability, random failures, environmental fluctuations | Missing data, measurement errors, uncertain parameters, model misspecifications, conflicting expert judgments |
| **Domains of Application** | Reliability engineering, quantitative risk assessment, probabilistic safety analysis | Early risk identification, strategic risk assessment, expert assessment in data-limited environments |
| **Examples** | Differences in component lifetime, variability in patient inflow | Unknown failure rate of new technology, divergence in expert estimates |
| **Reduction Strategy** | Managed through robustness, redundancy or design margins | Reduced through data collection, model validation and knowledge improvement |
| **Interrelation** | Provides baseline stochastic variability | Adds cognitive and informational uncertainty on top of aleatory variation |

## 3. Methods to Capture Uncertainty in Risk Management

To ensure analytical consistency across methodological categories, each approach is examined following a standardized structure. For every identified method, the general principle and theoretical foundation are introduced, representative models and applications are outlined, strengths and weaknesses in representing uncertainty are discussed, limitations are addressed and exemplary studies or domains of application are highlighted. The comparative position of these methods is further summarized in Table 2 and Table 3, which provide an overview of their core characteristics and their capacities to capture, quantify and visualize uncertainty.

Table 2
Overview of methodological approaches for capturing and representing uncertainty in risk management

| Category | Focus of the Method | Typical Techniques / Models | Type of Uncertainty Addressed | Strengths | Limitations |
|---|---|---|---|---|---|
| **Probabilistic Approaches** | Quantitative modeling of randomness, variability and stochastic dependence between risk factors | Probabilistic Risk Assessment (PRA), Bayesian Networks, Monte Carlo Simulation, Fault Tree, Event Tree Analysis | Primarily aleatory uncertainty (stochastic variability), limited handling of epistemic uncertainty through Bayesian updating and robust (minmax) decisions | Mathematically rigorous and transparent, allows quantitative comparison and sensitivity analysis, broadly standardized and accepted in engineering and safety-critical sectors | Requires extensive and reliable data, complex to use for ambiguous, incomplete or subjective information, assumes measurable probabilities even when knowledge is limited |
| **Evidence- and Fuzzy-Logic Models** | Representation of vagueness, imprecision and incomplete knowledge using linguistic variables or belief structures | Fuzzy Logic, Dempster-Shafer Theory, Possibility Theory | Primarily epistemic uncertainty, captures imprecise and subjective expert knowledge | Capable of representing uncertainty without precise probability distributions, flexible under incomplete information, useful for integrating expert judgment | High subjectivity in parameterization, lacks universal calibration standards, validation and interpretation can be challenging |
| **Qualitative Elicitation Methods** | Structured elicitation and synthesis of expert judgments under uncertainty, focuses on interpretation and consensus building rather than quantification | Delphi Method, Expert Judgment, Scenario Analysis, Risk Workshops | Primarily epistemic uncertainty, often combined with contextual and cognitive uncertainty | Enables assessment in data-scarce contexts, encourages interdisciplinary reflection, supports transparent reasoning and decision legitimacy | Strongly dependent on expert selection and framing, low reproducibility, results may be influenced by group dynamics and bias |
| **Graphical and Visual Methods** | Visual representation and communication of uncertain relationships, causal pathways or risk magnitudes, supports shared understanding | Bow-Tie Diagram, Influence Diagram, Risk Matrix, Uncertainty Map | Primarily epistemic uncertainty, with qualitative representations of aleatory aspects, typically without explicit distinction or formal modeling | Facilitates intuitive understanding and risk dialogue, enhances transparency and communication between disciplines, suitable for decision support in practice | Limited analytical depth, often qualitative, may oversimplify uncertainty, lacks formal quantification or integration with probabilistic data |
| **Hybrid and Integrated Approaches** | Combination of probabilistic, fuzzy and qualitative paradigms to achieve a more holistic treatment of uncertainty | Bayesian-Fuzzy Hybrid Models, Multi-Criteria Decision Analysis with Uncertainty Analysis, Integrated Risk Frameworks, Game-theoretic decision models as hybrid extensions that integrate strategic behavior with probabilistic or fuzzy representations | Aleatory and epistemic uncertainty and in some frameworks strategic or behavioral uncertainty, integrated across layers | Offers a comprehensive and system-oriented view, balances analytical rigor and interpretability, suitable for complex, multi-criteria decisions | Conceptually and computationally complex, resource-intensive, may be difficult to implement and communicate in operational settings |

## 3.1. Probabilistic Approaches

Probabilistic methods represent the most established group of techniques for capturing uncertainty in risk management [39–42]. They are based on the premise that uncertainty can be represented through probability distributions that describe the range of possible outcomes. This approach enables the derivation of statistical measures such as expected values, confidence intervals and quantitative risk indicators, including failure probabilities and value at risk. In this context, the Monte Carlo simulation is frequently used to analyze how uncertainty affects complex systems. This is achieved by repeatedly drawing samples from specified probability distributions for the input variables and evaluating the resulting variability in system outputs [43]. Bayesian networks integrate expert knowledge with observed data and update probabilities as soon as new information becomes available [44,45]. Classical techniques such as fault tree and event tree analysis additionally support the systematic representation of causal pathways that may lead to undesirable events [46]. These methods are comprehensively documented in key works on probabilistic risk assessment and reliability analysis and are firmly established in technical standards and safety regulations [47–49]. The application of probabilistic methods is particularly appropriate when uncertainty is largely driven by random variability and thus reflects aleatory uncertainty [42,50]. In situations where sufficient empirical data are available, stochastic behavior can be reliably characterized and probabilistic models provide robust and reproducible estimates. One reason probabilistic approaches are widely used is that their mathematical structure is explicit and can be systematically verified. Assumptions and model structures can be presented transparently and the resulting analyses are understandable and verifiable for independent experts. This has led to the widespread use of probabilistic risk assessment in safety critical sectors such as nuclear energy, transportation and clinical infrastructure, where formal quantitative indicators are required both for regulatory purposes and for technical verification [51,52]. The ability of probabilistic methods to represent uncertainty reaches its limits when uncertainty arises mainly from incomplete knowledge rather than randomness. In situations characterized by epistemic uncertainty, including limited data availability, sparse evidence, model ambiguity or divergent expert assessments, assigning precise probability distributions can create a misleading impression of objectivity [53,54]. In such cases, probabilities may inadequately capture the actual extent of uncertainty and suggest a level of accuracy that is not warranted given an incomplete or contested knowledge base. Another limitation concerns the structural assumptions of probabilistic models. Independence assumptions may conceal existing dependencies and assumptions regarding stationary or fixed distributional forms often fail to reflect the behavior of real-world systems. Such simplifications risk obscuring model uncertainty and may lead analysts to overestimate the reliability of numerical results. The apparent precision of probabilistic estimates may then be misinterpreted as accuracy and epistemic uncertainty may be underestimated or conflated with variability. These challenges have motivated a range of extensions to classical probabilistic risk analysis. Bayesian approaches have been refined to combine empirical data systematically with expert judgment and to represent parameter uncertainty through hierarchical structures and posterior distributions [44,45]. Robust optimization and risk-adaptive stochastic optimization address distributional uncertainty by evaluating not a single model but a range of plausible distributional assumptions [7]. In reliability engineering and system safety, probabilistic frameworks are increasingly combined with uncertainty quantification techniques to analyze how sensitive model outcomes are to assumptions about parameters, model structures and data quality. Recent review articles on decision making under uncertainty and stochastic optimization [7,55–58] emphasize that probabilistic approaches are evolving toward more flexible and integrative modeling strategies. They increasingly address high dimensional systems, context-dependent information and dynamic decision processes that have become highly relevant in areas such as critical infrastructures and healthcare [59–61]. Overall, probabilistic approaches remain central to quantitative risk assessment. They enable the quantitative treatment of aleatory uncertainty, support systematic sensitivity analyses and integrate well with formal decision-making tools. At the same time, their application requires careful reflection on data availability, data quality [62], modeling assumptions and the distinction between aleatory and epistemic uncertainty. In contexts with limited or contested knowledge, probabilistic models benefit from being combined with approaches that represent vagueness, ambiguity and ignorance more explicitly. Recent research has increasingly expanded and refined probabilistic methods in risk assessment. Classical modeling structures are being extended and adapted to increasingly complex, data-rich and dynamic environments. Important developments include modular and scenario-adaptive architectures within probabilistic risk assessment, refined Bayesian networks for high-dimensional systems and the growing integration of probabilistic reasoning with machine learning methods, which improves both predictive performance and uncertainty quantification [59,63]. A consistent trend in recent research is the enhanced treatment of structured epistemic uncertainty, for example through hierarchical Bayesian models, robust distributional assumptions or advanced uncertainty quantification techniques [64,65]. These developments illustrate the central role of probabilistic approaches and their ongoing theoretical evolution, driven by increasing system complexity, technological innovation and the availability of heterogeneous data sources. However, these approaches typically rely on assumptions regarding data availability and model structure, which may not hold under conditions of severe epistemic uncertainty.

## 3.2. Evidence and Fuzzy Methods

Evidence-based and fuzzy methods are useful when uncertainty is not adequately expressible through precise probabilities [66,67]. These approaches address the restrictive assumptions of probabilistic models and enable vague, incomplete or conflicting information to be represented in a structured and mathematically consistent way [68–70]. The underlying principle of these methods is to capture imprecision and partial belief rather than to model stochastic randomness. Rather than assigning exact probabilities, these methods represent uncertainty using membership functions, belief and plausibility measures or possibility distributions [71,72], which reflect the quality and completeness of the underlying knowledge. Fuzzy logic, originally introduced by Zadeh [66], models uncertainty through the use of fuzzy sets that represent linguistic categories such as “approximately high” or “moderately likely”. This concept allows gradual transitions between states rather than binary classifications and facilitates reasoning in situations characterized by vague or qualitative information. The Dempster-Shafer theory [68], also known as evidence theory, extends classical probability by distinguishing between belief, plausibility and ignorance, thereby providing an explicit representation of incomplete or conflicting information. Possibility theory, which is closely related to fuzzy logic, quantifies the plausibility of outcomes under partial knowledge and uses possibility and necessity measures as complements to traditional probability [71,72]. A key strength of these methods lies in their ability to represent expert judgments and subjective assessments in a structured and mathematically consistent way [73,74]. These methods are particularly useful in contexts characterized by epistemic uncertainty, such as early risk identification, medical diagnostics or multicriteria decision analysis, where empirical data are scarce, ambiguous or unreliable [20,75]. By avoiding the need for precise probability assignments, these methods enable forms of uncertainty representation that correspond more closely to the ways in which experts reason qualitatively about ambiguity, conflict or incomplete knowledge [72,76]. Despite these advantages, evidence-based and fuzzy methods also have limitations, as their parametrization often depends heavily on expert judgments, which can reduce reproducibility and introduce bias [73,74]. Membership degrees, belief functions and possibility measures may be less intuitive for practitioners to empirically substantiate than probabilistic metrics and the aggregation of evidential or fuzzy information across sources or models can lead to additional ambiguity or increased computational complexity [20]. These limitations have constrained the use of such methods in regulated, data-driven domains of risk management that require standardized quantitative outputs [75]. Overall, evidence-based and fuzzy methods offer a valuable means of representing uncertainty in contexts dominated not by random variability but by epistemic uncertainty [77]. They complement probabilistic approaches by explicitly capturing vagueness, partial belief and ignorance, thereby strengthening the theoretical basis for representing uncertainty in risk analysis. Foundational work on fuzzy sets, evidence theory and possibility theory provides the conceptual basis for their use in risk assessment and decision support [66,68,72,76]. Current research continues to refine and expand the application of evidence-based and fuzzy methods in risk analysis. Ongoing developments focus on improving the aggregation of conflicting evidence [78–81], enhancing the interpretability of belief and membership functions through new uncertainty measures [82–84] and integrating evidential or fuzzy approaches with probabilistic or data-driven methods [6,85,86]. These advances reflect a growing interest in hybrid formulations that combine fuzzy sets, belief structures or possibility measures with machine learning, Bayesian updating or multicriteria decision models. Across recent studies, the shared objective is to strengthen the representation of epistemic uncertainty in complex systems and to address challenges arising from limited data, ambiguous information and expert disagreement [87–89]. This evolving field reflects the growing importance and development of evidence-based and fuzzy approaches in risk assessment. While these methods provide greater flexibility in representing incomplete or imprecise information, they often face challenges related to parameterization, interpretability, and standardization across applications.

## 3.3. Qualitative Elicitation Techniques

Qualitative elicitation techniques aim to systematically capture expert knowledge, interpretative reasoning and context-specific insights when empirical data are insufficient, uncertain or unavailable [90–92]. Instead of expressing uncertainty through numerical measures, these approaches analyse how professionals interpret uncertain information and assess it in relation to their disciplinary and situational knowledge [74,93,94]. The central assumption underlying qualitative elicitation techniques is that human judgment and collective expertise represent a key source of evidence, particularly in situations of epistemic uncertainty where system behavior cannot be reliably determined exclusively on the basis of quantitative data [90,95,96]. Qualitative approaches structure expert knowledge in risk assessment across a wide range of techniques. These approaches include structured interviews, Delphi methods, focus groups, cognitive mapping procedures, scenario analyses and facilitated risk workshops [97–101]. Such techniques support experts in making assumptions explicit, identifying interdependencies and examining the implications of uncertainties in complex socio technical systems. Cognitive maps and scenario-based approaches make it possible to reveal causal mechanisms, emerging risks and context specific dependencies that are often underrepresented in quantitative models [102–104]. Delphi methods and iterative consensus-building processes reduce the influence of individual biases and enable collective expert judgments to evolve through controlled feedback [98,105–107]. The main advantage of qualitative elicitation is its ability to make implicit knowledge, divergent interpretations and context specific factors visible, which can only be captured to a limited extent by probabilistic or formal mathematical approaches [92,108,109]. These techniques increase the transparency of risk assessments because they systematically make the reasoning behind expert judgments, along with the underlying assumptions, value considerations and confidence levels, explicit [110,111]. This is particularly important in complex projects characterized by distributed and sometimes conflicting knowledge, high levels of uncertainty, multidisciplinary decision making and systemic interactions that are only partially quantifiable [112,113]. Qualitative elicitation is therefore especially relevant for early risk identification, strategic decision making and the assessment of emerging risks, including those arising from new technologies where robust empirical evidence is not yet available. The participatory nature of qualitative elicitation supports organizational learning and strengthens the acceptance of risk management outcomes. Despite these advantages, qualitative elicitation techniques exhibit important limitations, because their results depend heavily on expert selection, elicitation design and the quality of facilitation [91,108]. Group dynamics, anchoring, framing effects and other cognitive biases may influence judgments and differences in expertise or dominance can shape discussions and conclusions [95,114]. In addition, the resulting assessments are often difficult to verify or compare systematically and the lack of formal metrics reduces their reproducibility relative to quantitative approaches [115]. To mitigate these limitations, elicitation procedures are increasingly combined with semi- quantitative rating schemes, structured decision analysis or probabilistic models [116,117]. In summary, qualitative elicitation methods are essential in situations where empirical evidence is limited and expert judgment constitutes a key source of information. They complement quantitative and formal approaches by capturing interpretative and context-related dimensions of uncertainty and by providing a more comprehensive basis for risk related decision making [74,94,118]. Their theoretical foundations build on research in expert judgment, decision making and interpretive risk analysis, which provides a basis for their use in risk management [109]. Current research increasingly focuses on enhancing procedural transparency, reducing cognitive biases and integrating interactive or digital environments that support group-based reasoning [99,101,119]. Hybrid approaches that integrate qualitative elicitation with Bayesian updating, machine assisted interpretation, multicriteria decision frameworks or other integrative methods are attracting increasing attention [106,116,120,121]. These developments indicate an increasing effort to integrate expert-derived insights systematically into uncertainty analysis and to improve the traceability of qualitative assessments in complex systems. Recent contributions demonstrate that qualitative elicitation remains a central component in the representation of uncertainty [122].

### 3.4. Graphical and Visual Representations

Graphical and visual methods shift the emphasis from numerical quantification toward the representation and communication of uncertainty through visual means [123–125]. Their underlying assumption is that complex relationships, dependencies and uncertainties can often be interpreted more intuitively using graphical structures than numerical tables or analytical expressions [126–128]. Such approaches make it easier to identify the location, magnitude and structural characteristics of uncertainty, including its distribution across system elements and its interdependencies [129–131]. Typical graphical representations of uncertainty include influence diagrams, causal diagrams, bow-tie diagrams, risk maps and network-based visualizations [132–140]. Quantitative representations of uncertainty are often illustrated using probability density functions, fuzzy membership curves or uncertainty bands [124,127,141]. More advanced visualization approaches [129,142–145] rely on multidimensional visualization tools, interactive dashboards and scenario-based exploration environments that visually overlay uncertainty in terms of both magnitude and the reliability of underlying estimates, thereby enabling dynamic exploration. Graphical and visual representations support decision making by rendering abstract or complex uncertainty structures accessible and by revealing critical dependencies, systemic vulnerabilities, emerging risks and existing knowledge gaps that are not readily captured by purely analytical models [127,128,146]. Their intuitive nature facilitates communication between experts and non-experts, supports the development of shared mental models, understood as a common understanding of key system structures and relationships and enhances interdisciplinary discussion [128,130]. These characteristics highlight the potential role of graphical methods as complementary tools in safety-critical contexts and high-risk environments, where clear communication of uncertainty is essential. Despite their strengths, graphical methods exhibit important limitations: They are primarily descriptive and do not provide an independent quantification of uncertainty (with a few exceptions [147,148]). Such representations indicate where uncertainty arises within a system and how it is structured, but do not generate numerical probabilities or calculated risk indicators. Their interpretative nature may introduce subjectivity and visual elements can be misinterpreted if they are not designed or standardized with sufficient care [123,126,149]. The interpretation of visual representations of uncertainty is subject to cognitive biases, which may cause the same visualization to be interpreted differently by users with varying levels of experience and expertise [131,150]. Graphical approaches typically do not explicitly distinguish between aleatory and epistemic uncertainty unless combined with complementary modeling frameworks, such as probabilistic models including Bayesian formulations or evidential approaches, which allow a formal separation of different uncertainty types [137,138]. Visual representations of complex systems may result in oversimplification when the complexity of the underlying information cannot be adequately conveyed through graphical representation [123,126,151]. In this context, oversimplification refers to situations in which complex or multilayered uncertainty structures are reduced to an extent that essential details, dependencies or different forms of uncertainty are no longer adequately represented. In summary, graphical and visual methods provide an essential complement to analytical and probabilistic approaches by translating uncertainty into a form that supports interpretation, communication and joint decision making. They can enrich risk assessments by revealing structural relationships, highlighting knowledge gaps and supporting participatory evaluation processes. Research in visualization, decision support and cognitive psychology suggests that graphical representations can facilitate the understanding, transparency and interpretability of uncertainty in risk analysis [124,130,146]. Recent studies [129,136,142,143], [152] build upon the foundations of graphical uncertainty representations, with an increasing focus on interactive and dynamic visualization, integration with computational and probabilistic models and improved interpretability and standardization in participatory decision contexts. These developments underscore the increasing relevance of graphical and visual approaches to uncertainty communication in modern risk analysis. Although these approaches enhance transparency and stakeholder communication, they may lack the formal rigor required for quantitative decision-making or regulatory contexts.

### 3.5. Hybrid Approaches

Hybrid and integrated approaches are designed to combine the strengths of probabilistic, evidential, fuzzy, qualitative and graphical methods within unified frameworks for uncertainty assessment [73,153–155]. Hybrid approaches combine different paradigms, whereas integrated approaches embed these paradigms within a coherent analytical framework in which different forms of uncertainty can be represented simultaneously [156–158]. Their underlying premise is that no single method can fully capture the multifaceted nature of uncertainty. This includes aleatory variability, epistemic incompleteness and in certain contexts, strategic or behavioral uncertainty [25,159,160]. By integrating different forms of uncertainty representation, hybrid approaches aim to support a more comprehensive and coherent understanding of risks and their interdependencies in complex systems [155,161]. Hybrid models can take a variety of forms. They integrate probabilistic reasoning with approaches designed to address vague or incomplete information, aiming to capture both randomness and ambiguity within a single framework [162–166]. Other approaches link Monte Carlo simulations with evidence theory or combine Bayesian updating with qualitative elicitation to embed expert judgment in computational models [167,168]. Multi criteria decision frameworks frequently integrate quantitative performance indicators with linguistic assessments, thereby enabling the evaluation of alternatives when data quality varies or subjective interpretation is required [169–171]. Another category of hybrid approaches includes graphical procedures for specifying uncertainty, where experts visually delineate uncertainty regions by drawing rectangles that separately represent possible ranges of values for likelihood and impact [129,147,172,173]. These procedures combine visual representation with qualitative elicitation and allow experts to express ambiguity, confidence intervals and knowledge gaps without specifying precise numerical probabilities [174]. Such graphical elicitation formats support a structured interpretation of epistemic uncertainty and provide a cognitively accessible interface for interdisciplinary assessment teams, particularly in complex risk assessments in which uncertainties arise across different system levels, such as the technical level (for example sensor or equipment data), the process level (workflows and routines), the organizational level (structures and responsibilities), the human level (behavior and expertise) and the environmental or contextual level (regulatory or infrastructural conditions) [175]. Hybrid extensions increasingly incorporate advanced statistical and decision analytic components to capture complex dependencies and behavior related sources of uncertainty [120,159]. Hybrid Bayesian and evidential models are particularly suited to representing uncertainty arising from interactions between actors or subsystems that extend beyond purely aleatory or epistemic formulations [176,177]. When combined with probabilistic, fuzzy based or qualitative inputs, these models support decision making in domains such as security, resource allocation or the protection of critical infrastructures, where risk outcomes depend on the configuration and responses of multiple interacting agents or system components [178,179]. Hybrid models aim to link different forms of uncertainty in a consistent manner and to represent quantitative and interpretative elements as required for the respective application [153,154]. A central strength of hybrid frameworks is their flexibility, which enables heterogeneous uncertainties to be integrated within a single analytical framework. By capturing both variability and knowledge gaps, this integration reduces the risk of systematic epistemic limitations associated with single method approaches and increases robustness [155,180]. Beyond epistemic uncertainty, hybrid formats also address interpretative uncertainty arising from meaning, framing, stakeholder values or incomplete conceptual alignment [170,181]. As a result, such frameworks are particularly valuable for interdisciplinary risk assessments and for sociotechnical systems that integrate diverse information sources including quantitative data, model-based outputs and qualitative expert judgments [171,182]. At the same time, hybrid approaches introduce important theoretical and practical challenges. Integrating heterogeneous forms of uncertainty requires careful calibration, harmonization of scales and explicit assumptions about how different uncertainty types should be combined [157,159]. Increasing model complexity can reduce interpretability, while computational demands may rise substantially, especially in high dimensional or dynamic systems [120,161]. Validation also becomes more demanding, as multiple model components must be assessed jointly to ensure internal consistency and reliability [168,183]. In summary, hybrid and integrated approaches represent a significant development in uncertainty modeling because they enable different uncertainty forms to be combined within a shared analytical architecture [25,73,153]. This allows for a more comprehensive representation of aleatory and epistemic uncertainty and, where relevant, strategic uncertainty. Foundational work on hybrid modeling, integrated risk frameworks and multi-criteria decision analysis provides theoretical justification for their application in risk analysis [156,184]. Current research continues to expand the theoretical basis of hybrid approaches by combining probabilistic models with fuzzy or evidential reasoning, embedding expert elicitation within computational frameworks and linking hybrid structures to Bayesian networks or uncertainty quantification techniques [162,163,185]. Recent research also addresses coupled probabilistic and deterministic analyses, as well as the integration of machine learning with fuzzy-based or evidential representations to support explainability [186,187]. Against this background, hybrid approaches are increasingly viewed as an integrative component of modern risk analysis, particularly in complex environments where multiple types of uncertainty must be considered simultaneously [180,187]. Rather than replacing established methodologies, hybrid approaches provide a unifying perspective that emphasizes coherence across methods.

Table 3
Comparison of methodological approaches by their ability to capture, quantify and visualize uncertainty

| Method Category | Capture | Quantify | Visualize |
|---|---|---|---|
| **Probabilistic Approaches** | •• (through stochastic modeling of random variables) | • • • (fully quantitative based on probabilistic modeling) | • (not primarily visual) |
| **Evidence- and Fuzzy-Logic Models** | •• (represent vague or incomplete knowledge) | • (semi-quantitative) | • (limited graphical representation possible) |
| **Qualitative Elicitation Methods** | ••• (strong in eliciting and structuring expert knowledge) | X (non-quantitative) | • (scenario or workshop-based visualization) |
| **Graphical and Visual Methods** | • (implicitly capture structure) | X (non-numerical) | • • • (primary focus) |
| **Hybrid and Integrated Approaches** | •• (combine multiple methods to capture different uncertainty aspects) | •• (partial quantification depending on included quantitative components) | • (depends on integration of complementary methods and may support visualization through combined graphical and analytical components) |

Note: X not applicable, • low relevance or coverage (limited or indirect support for the criterion), •• moderate relevance or coverage (partial or context-dependent support), ••• high relevance or coverage (explicit and systematic support as a core feature of the methods)

## 4. Comparative Analysis and Discussion

We found no single approach to sufficiently address all dimensions of uncertainty, which suggests the need for context-dependent method selection. Across the methodological families, each approach addresses only specific facets of uncertainty. Probabilistic methods are most appropriate when uncertainty is predominantly aleatory and when sufficient data are available to support the specification of probability distributions and the calibration of quantitative models. Evidence and fuzzy logic approaches are particularly suited to contexts in which knowledge is incomplete, information is vague or expert judgment plays a central role. Qualitative elicitation techniques and graphical representations are especially useful in settings with complex problem structures, heterogeneous stakeholders or decision processes that require explicit documentation of assumptions and reasoning. From a comparative perspective, the choice of method largely depends on three factors. The first is the dominant type of uncertainty, in particular the balance between aleatory and epistemic components. The second is the availability, quality and stability of data, which constrains the extent to which quantitative calibration is feasible. The third is the decision context, including regulatory requirements, time constraints and the need for interdisciplinary or participatory processes. Hybrid and integrated approaches respond to these conditions by combining methods to capture different uncertainty dimensions within a single analytical architecture. However, their advantages in terms of coverage and flexibility must be weighed against increased complexity and higher demands regarding transparency, validation and communication.

## 5. Operationalizing the Findings in Practice

This section operationalizes the comparative findings by outlining a structured decision pathway for selecting uncertainty representation methods in risk management. The pathway is structured around three primary selection drivers: (i) dominant uncertainty type, (ii) data availability and quality, and (iii) decision context and required outputs. Based on these drivers and the reviewed literature, five advisory steps are proposed to support transparent and context-appropriate method selection in research and practice. To facilitate justification of the selected method, a checklist-based documentation approach is outlined in the Appendix. The proposed pathway is intended as an evidence-informed synthesis and documentation aid rather than an empirically validated decision model. While presented sequentially, the steps should be understood as an iterative decision guidance. In practice, data constraints and decision requirements may require revisiting earlier steps. Figure 1 presents the proposed decision pathway for selecting uncertainty representation methods in applied risk management.

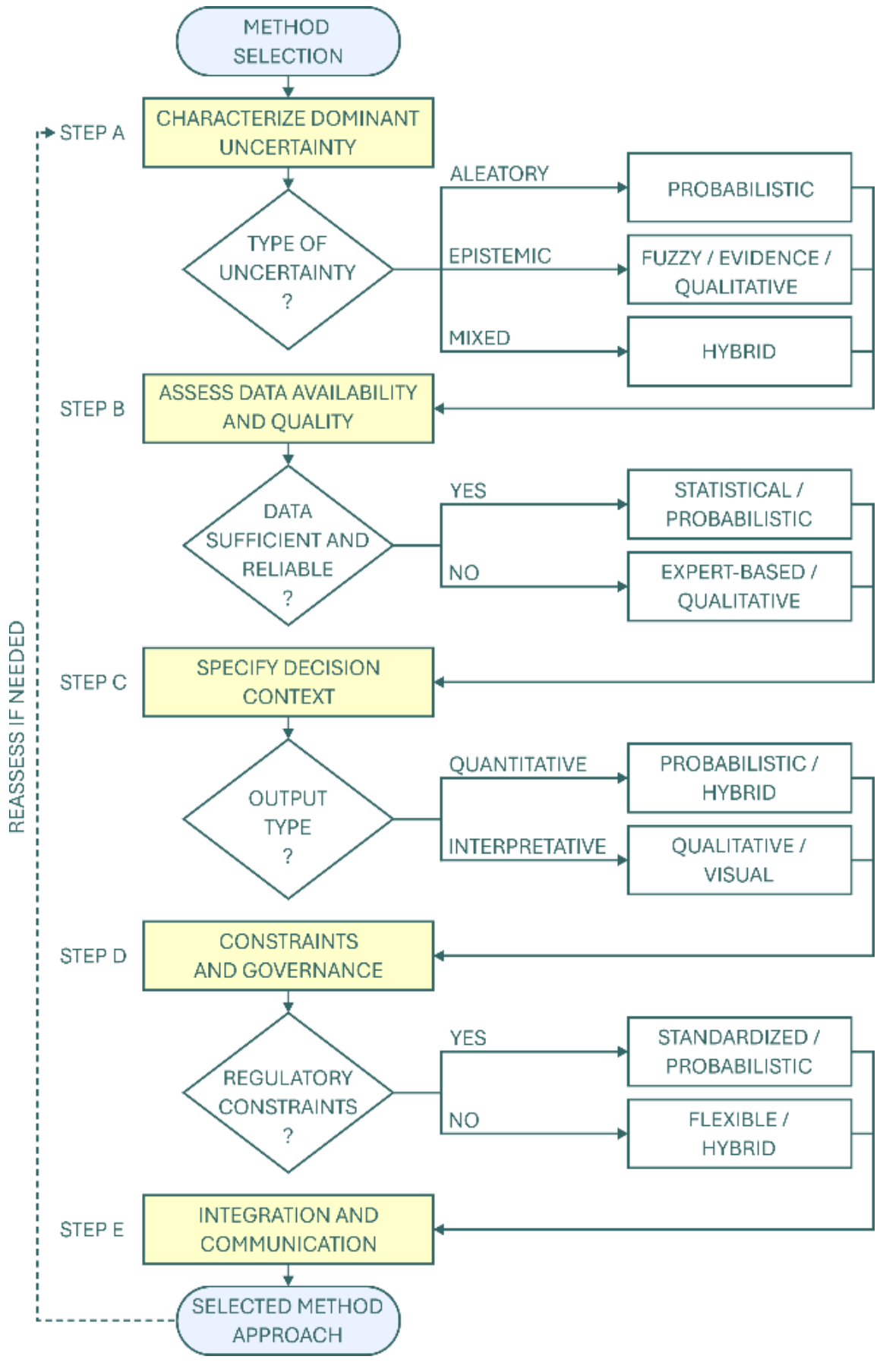

Figure 1. Structured decision pathway for selecting uncertainty representation methods in applied risk management.

The proposed pathway is conceptual and synthesizes findings from the reviewed literature. It is intended as structured guidance rather than as an empirically validated predictive model and therefore requires further testing in domain-specific applications. From an epistemological perspective, uncertainty may be interpreted either as inherent variability or as a limitation of knowledge. In practice, however, this distinction is not always clear-cut, since uncertainty types may overlap or be interpreted differently depending on context and available information. In particular, ambiguity and incomplete information may make it difficult to assign uncertainty unambiguously to a single category. Therefore, the selection of methods should not be understood as a strictly deterministic process but as a context-dependent decision that requires informed and context-sensitive judgment. The pathway offers a systematic orientation for method selection, but it does not replace the need for critical assessment of the underlying assumptions about uncertainty and its representation.

## 5.1. Characterize the dominant uncertainty

Method selection should begin with an explicit characterization of the dominant uncertainty type. Distinguishing between aleatory and epistemic components is essential because the suitability of probabilistic, evidential, qualitative, graphical, or hybrid methods depends on the underlying uncertainty profile.

- If uncertainty is predominantly aleatory (Section 2.2.1), meaning it reflects stochastic variability in stable and repeatable mechanisms, probabilistic approaches (e.g., PRA, Bayesian networks, Monte Carlo simulation) are most appropriate for quantitative estimation and sensitivity analysis.

- If uncertainty is predominantly epistemic (Section 2.2.2), meaning it arises from incomplete, imprecise, or contested knowledge, evidence-based and fuzzy-logic models (belief functions, possibility theory, linguistic variables) provide a suitable representation without requiring precise probability assignments.

- If uncertainty is mixed or layered, combining aleatory and epistemic components or including strategic and behavioral aspects across multiple system levels, hybrid or integrated approaches (Section 3.5) are suitable to represent multiple uncertainty forms coherently within a single analytical framework.

## 5.2. Assess data availability and quality

The second step involves assessing the availability and quality of the data used for uncertainty representation. Importantly, numerical precision should not be conflated with accuracy, since highly precise numerical outputs may still reflect substantial epistemic uncertainty. To avoid spurious precision, assumptions and confidence levels should be documented explicitly. Prior work highlights that an "illusion of accuracy" can arise when numerical results appear overly precise without adequate evidential support [188].

- If data are sufficient and reliable to support the estimation of statistical distributions, probabilistic models (Section 3.1) are applicable. However, sensitivity analyses and robustness checks should be included, and key modeling assumptions (e.g., independence, stationarity, distributional form) should be made explicit.

- If data are sparse, heterogeneous, or of limited reliability, and uncertainty relies substantially on expert judgment, qualitative elicitation techniques (Section 3.3) are appropriate to structure assumptions and reasoning. In such contexts, evidence-based and fuzzy-logic models (Section 3.2) can complement elicitation by formally encoding imprecision and partial belief without requiring precise probability assignments.

- If interpretability and traceability are prioritized over numerical precision, graphical and visual methods (Section 3.4) are suitable for communicating uncertainty structure, dependencies, and knowledge gaps. Since visual representations are typically descriptive, they should be linked to analytical models when quantification or comparability is required.

## 5.3. Specify decision context and required outputs

The third step is to specify the decision context and the outputs required from the assessment. Method selection should align with the type of decision to be supported, the intended audience, and the required form of evidence. This includes whether the decision process requires quantitative indicators, comparative rankings, or interpretative and communicative outputs. In practice, the chosen uncertainty representation should match both the informational needs of stakeholders and potential requirements for traceability, transparency, and auditability.

- If the decision context requires quantified risk indicators, comparative metrics, or formal trade-off analyses, probabilistic models (Section 3.1) are suitable when uncertainties are largely aleatory and data permit statistical calibration. If epistemic components are substantial, hybrid approaches (Section 3.5) should be preferred to prevent spurious precision and to represent knowledge limitations explicitly.

- If the primary requirement is transparent reasoning, interpretability, and clear documentation of assumptions, qualitative elicitation techniques (Section 3.3) and graphical representations (Section 3.4) provide strong support. These methods make underlying assumptions and rationales explicit, facilitate interdisciplinary discussion, and enable communication of uncertainty in a form accessible to both experts and non-experts.

## 5.4. Consider constraints and governance requirements

The fourth step is to consider institutional constraints and governance requirements that shape method feasibility and acceptability. In many domains, risk assessment is conducted under regulatory expectations, organizational standards, or practical limitations regarding time, resources, and available expertise. These constraints influence not only the selection of methods but also the extent to which uncertainty representations can be justified and defended in formal decision processes. Explicit documentation of assumptions and uncertainty sources is therefore essential, particularly when decisions have safety-critical implications.

- In regulatory or engineering contexts where formal quantification and auditability are required, probabilistic approaches are often expected due to their established standards and formal structure. However, where epistemic uncertainty is material, results should be complemented by explicit uncertainty characterization, robustness analyses, and clear documentation of assumptions to reduce the risk of spurious precision.

- In multi-criteria decision contexts, uncertainty representation must account for ambiguity that arises from weighting, aggregation, and preference structures. If decision goals can be weighted meaningfully, scalarization approaches (i.e., aggregating multiple objectives into a weighted sum) are commonly applied and allow mathematically grounded trade-offs between competing goals [189]. If goals are strictly prioritized, lexicographic decision rules may be more appropriate, selecting solutions that optimize the most important objective first and using lower-priority objectives to break ties [190]. In both cases, uncertainty regarding weights, preferences, and stakeholder values should be treated explicitly, as it can substantially influence risk prioritization and decision outcomes.

## 5.5. Plan integration and communication

The fifth step is to plan how uncertainty representations will be integrated into decision processes and communicated to stakeholders. In applied contexts, method selection is not only a technical issue but also a communication and implementation challenge. Effective uncertainty treatment requires that results remain interpretable, that assumptions are traceable, and that uncertainty information can be conveyed in a form that supports decision-making. In particular, visualization can support understanding of where uncertainty arises, how large it is, and how reliable the underlying evidence is.

- If uncertainty representations are intended to support interdisciplinary decision processes, visual and graphical formats should be selected to communicate structure, dependencies, and evidence quality. Since visual outputs are typically descriptive, they should be linked to analytical models where quantitative defensibility or comparability is required.

- If expert judgment is a central information source, structured elicitation procedures should be applied to enhance transparency, mitigate cognitive biases, and document confidence levels. Where feasible, elicited knowledge should be connected to analytical models, for example through Bayesian updating or evidential formalisms, to support traceability and iterative revision as new information becomes available.

- If uncertainties are distributed across multiple system levels or involve mixed uncertainty types, a hybrid or integrated architecture should be specified. This includes explicit interfaces between model components, clear rules for combining uncertainty information, and documentation of assumptions to ensure internal coherence and interpretability.

## 5.6. Application Example

To illustrate the application of the proposed decision pathway, the following example demonstrates method selection in a practical risk assessment context. For example, consider a risk assessment in a healthcare setting within the context of critical infrastructure, such as the failure of a hospital power supply system resulting from a loss of external power supply as well as potential disturbances in internal backup, switching, and distribution systems, including possible delays or malfunctions in the activation of backup systems. In this scenario, uncertainty arises both from stochastic variability in system loads and component failure rates and from incomplete knowledge regarding functional dependencies and disturbance propagation mechanisms within the system. This includes interactions between the primary power supply, backup generators, fuel supply systems, as well as the dynamic load demands of critical and energy-intensive medical systems, technical building systems, and information technology infrastructure. Accordingly, the uncertainty can be characterized as mixed, comprising both aleatory and epistemic components. Data availability is limited to moderate. While historical data on component reliability and load profiles are available, significant uncertainties remain regarding functional system interactions, the performance of emergency measures, and the effectiveness of redundancy and backup configurations. These uncertainties are further exacerbated by the fact that hospital infrastructures are typically site-specific and differ substantially in terms of grid connection concepts, including potential high-voltage connections, as well as switching and transfer behavior. As a result, standardized data-driven assessment is only partially feasible, making the incorporation of expert judgment necessary. The decision context requires both quantitative indicators for operational planning and transparent communication of assumptions and uncertainties to interdisciplinary stakeholders, including technical personnel and hospital management. Following the proposed decision pathway, this combination of uncertainty characteristics, data conditions, and decision requirements suggests the use of a hybrid approach. In particular, the simultaneous consideration of stochastic variability and epistemic uncertainty necessitates the integration of different modeling approaches. Such an approach may combine probabilistic methods for modeling component failures and load variability with expert-based or evidence-based methods for assessing system interactions, switching processes, and emergency behavior. This enables the derivation of quantitative risk indicators while explicitly representing uncertainties related to system behavior and underlying assumptions. In addition, graphical representations can be employed to communicate the structure, dependencies, and implications of uncertainty in a transparent and comprehensible manner, particularly in interdisciplinary decision-making processes.

Overall, combining methods can be advantageous when a single paradigm cannot credibly capture the uncertainty profile (e.g., Bayesian updating of expert inputs or evidential layers for ambiguous parameters on top of probabilistic cores). Where data are limited or contested, transparency of reasoning and sensitivity analysis should be prioritized, and hybrid approaches may offer advantages over single-paradigm formulations. The next section discusses remaining research gaps, including the need for empirical evaluation of decision pathways and integration frameworks.

## 6. Research Gaps and Future Directions

The review identifies several gaps that warrant further research. Comparative empirical evaluations of methods remain rare, particularly studies that apply different approaches to the same risk problems and systematically analyze differences in outcomes and decision implications. The decision pathway and checklist proposed in Section 5 are conceptual contributions and future empirical evaluation in applied risk management contexts therefore remains an important direction for future research. Beyond these gaps, there is a noticeable lack of domain-specific applications in critical infrastructures and health care, where uncertainty is pronounced and choices of methods have direct implications for safety and service continuity.

Further challenges concern the operational integration of uncertainty methods into decision support systems, organizational procedures and digital tools. This includes questions regarding how uncertainty representations influence decision quality, stakeholder trust and the prioritization of risk control measures. Finally, several open questions remain regarding hybrid and visualization-based approaches, including the calibration and validation of combined models, the standardization and evaluation of uncertainty visualizations and the design of elicitation protocols that link qualitative insights with formal analytical structures. Addressing these gaps would support a more systematic and context-sensitive treatment of uncertainty in applied risk management.

## 7. Conclusions

This work examined how different methods represent uncertainty in risk management. The analysis demonstrates that uncertainty is conceptually multifaceted and that its treatment varies substantially across the literature. Probabilistic models provide a rigorous framework for quantifying aleatory uncertainty and continue to serve as the primary analytical basis of quantitative risk assessment. Evidence-based and fuzzy approaches build on this foundation by addressing vagueness, partial belief and incomplete knowledge and therefore offer a more explicit representation of epistemic uncertainty. Qualitative elicitation techniques, graphical representations and hybrid frameworks contribute interpretative and participatory dimensions that become essential when data are limited, information is inconsistent or incomplete and decisions involve multiple stakeholders. Across the reviewed literature, no single approach appears sufficient to cover the full range of uncertainty found in complex systems. These findings are consistent with prior work in risk analysis literature. Probabilistic techniques offer strong analytical capabilities but can produce misleading impressions of precision when applied to insufficiently constrained epistemic uncertainty. Qualitative and graphical approaches support interpretative depth and communication yet often lack the formal structure required for regulatory or design-oriented applications. Evidence-based and fuzzy methods provide greater flexibility in representing incomplete knowledge. However, they introduce challenges related to parameterization, interpretability and validation. Hybrid and integrated frameworks mitigate some of these limitations by combining quantitative and qualitative elements, although they also increase the practical complexity and require careful calibration and explicit assumptions to ensure coherence. The review points to two areas where additional theoretical progress would be beneficial. The first concerns conceptual clarification, in particular a more consistent distinction between aleatory and epistemic components and a clearer conceptual treatment of epistemic subcomponents, including ambiguity, model uncertainty and residual ignorance. The second concerns operational integration, namely the embedding of uncertainty methods into decision support processes, organizational practices and domain-specific applications such as critical infrastructures. Future research would benefit from comparative empirical evaluations of methods, systematic analyses of how uncertainty representations affect decision quality and further development of hybrid and visualization-based approaches that make uncertainty both analytically tractable and communicable to decision makers. Overall, the review suggests that representing uncertainty systematically is not a supplement to risk assessment but a central requirement for transparent, robust and defensible risk management. By synthesizing fragmented methodological traditions into a unified comparative framework, this review provides a foundation for more transparent and context-appropriate uncertainty handling in complex risk environments.

# Appendix A

To complement the selection guidance in Section 4, this Appendix proposes a checklist-based documentation format to support transparent justification of the selected uncertainty representation method. Table 4 provides documentation prompts to record key assumptions and decision criteria prior to committing to a specific approach.

Table 4
Checklist for documenting method selection prior to commitment

| Dimension | Key checks | Documentation prompts |
|---|---|---|
| **Uncertainty profiling** | Classification and localization of uncertainty sources | Have key uncertainties been classified as aleatory, epistemic, or mixed, and have their sources within the system been identified? |
| | Epistemic subcomponents relevant for method choice | Are epistemic subcomponents (e.g., parameter uncertainty, model uncertainty, ambiguity) material for representation or communication? Can the chosen method distinguish these components? In many cases, this requires hybrid approaches. |
| **Data fitness and assumptions** | Data adequacy for calibration | Do available data support statistical calibration (e.g., distributional estimation)? If not, have non-probabilistic complements (e.g., evidential, fuzzy, qualitative) been considered to avoid spurious precision? |
| | Model assumptions recorded | Have critical assumptions been documented (e.g., independence, stationarity, model form, boundary conditions)? |
| | Robustness of conclusions | Is a sensitivity or robustness analysis planned to test the influence of assumptions and modeling choices? |
| **Decision context and outputs** | Output requirements aligned with method choice | Which outputs are required (e.g., numerical indicators, ranges/confidence bands, narratives, visuals), for whom, and for which decisions? Is the selected method family aligned with these requirements (e.g., quantitative outputs: probabilistic/hybrid approaches; interpretive or participatory outputs: qualitative/visual methods)? |
| **Governance constraints and auditability** | Regulatory and documentation requirements | Are regulatory or audit requirements present that mandate traceability, defensibility, or formal quantification? If yes, does the chosen approach support reviewable documentation? |
| **Communication and integration** | Stakeholder-oriented communication | Are uncertainty representations and visualizations selected to match stakeholder needs and interpretive capabilities? |
| | Link between visuals and analytics | Where quantification is required, are visual outputs linked to underlying analytical models to support interpretability and defensibility? |

| | |
|---|---|
| Expert judgment traceability | If expert judgment is used, are elicitation procedures structured and transparent, and are uncertainty statements linked to documented assumptions and confidence levels? |
| Hybrid architecture specified | If uncertainty is layered across system levels or perspectives, is a hybrid/integrated architecture specified including interfaces and rules for combining uncertainty representations? |
| Updating and iteration planned | Is an updating strategy defined for incorporating new evidence and communicating how uncertainty assessments evolve over time? |

## Acknowledge

This work was supported by the LIT Secure and Correct Systems Lab funded by the State of Upper Austria and the Linz Institute of Technology (LIT-2019-7-INC-316).